\documentclass[11pt,twocolumn]{article}
\usepackage{graphicx}
\usepackage{enumitem}
\usepackage{fourier-orns} 
\setitemize{noitemsep,topsep=0pt,parsep=0pt,partopsep=0pt}

\begin{document}
 \twocolumn[
 \hrule 
 \vspace{0.5cm}
 {\bf 
 Scientific proposal submitted to the General Sciences committee for the 
 July-December 2012 running cycle at the Advanced Light Source at the Lawrence
 Berkeley National Laboratory. Experiment ID: ALS-05068. Score: 2.00 
 Beamline: 10.0.1.2 Shifts Requested: 40 Shifts Allocated: 15. 
  
 Contact information: Guillermo Hinojosa, hinojosa@icf.unam.mx, Instituto de
 Ciencias Físicas, Apartado Postal 48-3, Cuernavaca 62210,  Universidad Nacional
 Autónoma de México UNAM-Campus Morelos, México. } 
 
\vspace{-0.3cm}  
 
\begin{center} \leafNW  \end{center} 
 
 \hrule 
 
\vspace{1cm} 
 
 \begin{@twocolumnfalse}
 \begin{center}
\begin{center} {\footnotesize \bf
 PHOTOIONIZATION OF CHLORINE, SULFUR, PHOSPHORUS AND A HYPOTHESIS TO QUENCH METASTABLE STATES}
\end{center}  
 {\footnotesize by Guillermo Hinojosa}\\

 \end{center}
 
 The objective of our proposal is to study the photoionization of Chlorine, Sulfur, and Phosphorus ions 
 using the unique high resolution capability of 10.0.1 beamline in the ALS. These species are 
 relevant in astrophysics and as a benchmark for state of the art theoretical models of the quantum
 properties of ions. We also propose to test a technique to quench the metastable states from the 
 ion-beam based on recent experimental evidence and on a well known technique in chemistry. 
 The work of this proposal may lay the foundations for a definite solution to control the
 metastable component of the initial ion-beam.

 \begin{flushright} Request of 40 shifts.  \end{flushright}
  
 \end{@twocolumnfalse}]
 

 We propose to study ions that originate from open-shell atoms. They are of relevance
 because the electron correlation and  the inter channel coupling effects in these systems 
 are very strong and, therefore, they are a natural benchmark for detailed comparisons of 
 the LS and $jj$ coupling models\footnote{An essential question in quantum mechanics is to 
 compare the importance of the spin-spin, orbit-orbit and, spin-orbit interactions in middle-sized
 atoms quantum models.}. For this reason and with the unique high resolution features of the ALS\cite{cubaynes2004}, 
 data from the photoionization of these ions are expected to provide a great deal of details to compare with theory. 
 From a computational point of view, the main challenge in 
 the description of open shell systems is the inclusion of a large number of states. 
 For this reasons we expect that these data will encourage the development of theory to a 
 level that is becoming more and more available thanks to the continuous development 
 of computer power.
 
 In addition, accurate photoionization cross sections and high resolution spectra of ions of
 astrophysical interest are required to model a large variety of objects such as active galactic 
 nuclei, HII regions, planetary nebulae, novae, and supernovae.  Photoionization of ions of 
 Chlorine, Sulfur and Phosphorus have critical roles in interstellar chemistry. For instance, 
 they happen to be very reactive with H$_2$, forming hydrogenic molecular ions such 
 as ClH$_2$ which is the best tracer of optically thick H$_2$ components in diffuse 
 clouds \cite{Sonnentrucker2006}. 

 There is one single line of ClIII that is identified and used in abundance determination
 models \cite{witter2001}. In spectroscopy, a single line is rarely a definite proof of
 identification. We propose to measure, with high resolution, the absorption spectra
 of this particular species which is very important in astrophysical observations.
 
 Sulfur and its ionic states have  been discovered in the plasma of Jupiter's satellite Io and
 spectroscopic lines of its ions are routinely detected in the Jovian auroras \cite{morrissey1997}. 
 A number of strong features due to Sulfur ions have been observed in the spectra of the Io plasma
 torus, of the Sun and in stellar transition regions \cite{feldman2004}. Cross sections and precise
 spectroscopic lines measured with the high accuracy of beamline 10.0.1 would be of great aid in 
 the understanding of these environments and in the design of future spectrometer space satellites.
 
 To our knowledge, there exist high resolution data on the isoelectronic series of Chlorine-like 
 ions such as Ar$^+$ \cite{covington2011}, Ca$^{3+}$ \cite{ghassan2010}, Al$^{2+}$ \cite{aguilar2003}
 and Ar$^{5+}$ \cite{wang2007}. For the ions that we propose to investigate, just low resolution data 
 for S$^{+}$ \cite{kristensen2002} are available. For the case of Chlorine, only the pioneering
 work on the atomic case of Ru$\check{\mbox{s}}\check{\mbox{c}}$\'ic and Berkowitz has been reported
 \cite{rusctic1983}. 
 
 The only issue in terms of feasibility for the present proposal, is the production of a
 reliable beam of Chlorine. To show that it is possible to produce such an ion-beam, a Cl$^{+}$ 
 beam was generated in our local laboratory with the following procedure:  
 a small amount of Ferric Chloride (FeCl$_{3}$) was heated directly inside the chamber of a
 filament-type ion source. This compound is available commercially and is harmless.  
 The results are very promising and demonstrate that it is possible to produce a stable beam of 
 Cl$^{q+}$ by heating this compound with the insertion oven of the ECR ion source. There are 
 similar recipes to generate beams of S$^{q+}$ and P$^{q+}$ \cite{charge2011}. 
  
 In this technique, the main general problem for the measurement of the cross section is the presence 
 of an undetermined amount of metastable initial states in the ion-beam. Basically, the 
 initial state is composed of two electronic states that have different photoionization 
 probabilities (cross sections). One is the ground state and the second is a metastable 
 state that is created by collisions with electrons inside the ion 
 source\footnote{For instance, for all noble gases $^{2}$P$_{3/2}$ 
 is the ground state and $^{2}$P$_{1/2}$ is the metastable state (with the exception of He).}. 
 Actual cross section measurements are an indeterminate combination of the two states.
 As a result, theoretical models have been forced to include an empirical factor to fit the
 experiment, literally ruining any detailed conclusive comparisons with theory.
 
 An early attempt to circumvent this problem consisted in measuring the attenuation fraction
 of an O$^{+}$ beam in collisions with N$_{2}$ \cite{covington2001}. A recent effort consisted
 in combining the ion-trap and the merged-beam techniques \cite{bizau2011}. What we propose here 
 is a very simple method that has been overlooked and consists of producing a
 ground state ion-beam rather than normalizing to independent experimental measurements.
 In chemistry, it is well establish that resonant
 collisions\footnote{Thermal collisions with atoms or molecules with similar masses.} quench 
 efficiently the metastable component of ions \cite{ibuki1983}; it is a fundamental effect that 
 has to do with the way internal energy gets shared in collisions. 
 
\begin{figure}[ht!]
\begin{center}
\includegraphics[width=\columnwidth]{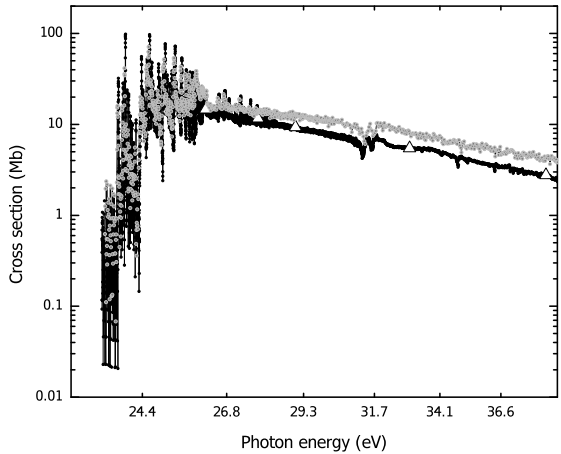}
\caption{\label{alsVSarhus} \footnotesize  Log-log plot of the photoionization of
 Kr$^{+}$. To show that our team has the capability to
 carry out this proposal and to illustrate the unique high resolution feature of the IPB in BL 
 10.0.1, we show data measured in the ALS by a team with the same team leader as that of
 the current proposal's group \cite{hinojosa2011}. The data consist in
 the cross sections measurements for the photoionization of Kr$^{+}$ (open triangles) 
 to which a broad photon energy scan taken at a resolution of 10 meV and steps of 1 meV has been
 normalized (black dots). To demonstrate that resonant collisions quench the metastable component, 
 the ALS' data are compared to pure ground state cross sections (gray dots) of Bizau {\it et al.} 
 \cite{bizau2011} (measured with lower resolution). When the metastable component is important, the
 cross section is expected to be higher, here both sets of data are comparable.}
\label{aparato}
\end{center}
\end{figure}

 First, we present two pieces of evidence to show that with the actual ion-photon beam (IPB) 
 end station in the ALS, it was possible\footnote{In its earliest stage the IPB had a filament-type 
 ion source.} to generate pure ground state beams when the ion-beam source was a hot-filament 
 source type that induced resonant collisions: (1) Cross section data on the photoionization 
 of Kr$^+$ \cite{hinojosa2011} turned out to be in agreement with similar data that was corrected 
 for the metastable contribution \cite{bizau2011} (see Fig. \ref{alsVSarhus}) and (2) The 
 photoionization of CO$^+$ \cite{hinojosa2002} cross section that was significantly 
 lower than that of a data set measured with an ECR ion source \cite{andersen2001}.
 
 In the ECR ion source, electrons couple with the source's radio frequency and collide with the 
 injected gas forming ions that remain in the plasma until they are expelled out of the ion source
 by electrostatic repulsion. While inside the ion source, these ions may suffer collisions with
 the surrounding gas. To test this idea we propose to inject a combination of Xe plus CH$_{4}$ 
 gases in the ECR to produce a Xe$^{+}$ ion-beam because it is well known that in this particular 
 collision system, the metastable state of Xe$^{+}$ is completely quench \cite{adams1980,Kok2000}. 
 The ECR would be operated at its lower power and at its higher pressure to enhance this effect. 
 Similar "quenching" collision systems for other ions (including Cl, S and P ions) are tabulated 
 in reference \cite{velazco1978}.

 In conclusion, we propose to measure systems that are {\it per se} relevant in atomic physics
 and in astrophysics. Because of its open shell character, these measurements will take full
 advantage of the high resolution power of the ALS. In addition, we will test a simple-to-check 
 method to control the metastable component of the ion-beam.
 
 The present proposal would be carried out by a team of scientist from the Institute of
 Physical Sciences at the National University of Mexico. This is a multidisciplinary group
 with familiarity in the use of synchrotron radiation. Some of the members are
 experienced in the ion-photon end station in beam line 10.0.1:

 \vspace{1.0cm}
 
 [List of co-authors has been removed by the author.]  
 
 \vspace{4.0cm} 
{\scriptsize 
\bibliography{biblio}
}

\begin{center} {\footnotesize \bf
 ACKNOWLEDGMENTS}
\end{center}  
Grant: DGAPA-IN113010

\vspace{2.5cm}

\begin{center}
{\Huge \leafSW}
\end{center}

\end{document}